\documentstyle[twocolumn,epsfig,prb,aps]{revtex}
\begin{document}
\draft
\title{$^{11}B$ NMR and Relaxation in $MgB_2$ Superconductor}
\author{J. K. Jung, Seung Ho Baek, F. Borsa\thanks{Also at Dipartimento
di Fisica and INFM, Universit\`{a} di Pavia, Pavia, Italy },
S. L. Bud'ko,
G. Lapertot,\thanks {On 
leave from Commissariat a l'Energie Atomique, DRFMC-SPSMS, 
38054 Grenoble, France} and P. C. Canfield}
\address{Ames Laboratory and Department of Physics and Astronomy, 
Iowa State University, Ames, IA 50011}  

\date{\today}
\maketitle
\begin{abstract}
$^{11}B$ NMR and nuclear spin-lattice relaxation rate (NSLR) are reported
at 7.2 Tesla and
1.4 Tesla in  powder samples of the intermetallic compound $MgB_2$ with
superconducting
transition temperature in zero field $T_c$ = 39.2 K. From the first
order quadrupole perturbed
NMR specrum a quadrupole coupling frequency of 835 $\pm$ 5 kHz is
obtained. The Knight
shift is very small and it decreases to zero in the superconducting
phase. The NSLR
follows a linear law with $T_1T$ = 165 $\pm$ 10 (sec K) . The results
in the normal phase indicate
a negligible $s$-character of the wave function of the conduction
electrons at the Fermi
level. Below $T_c$ the NSLR  is strongly field dependent indicating
the presence of an
important contribution related to the density and the thermal motion of
flux lines. No
coherence peak is observed at the lower field investigated (1.4 T).
\end{abstract}
\pacs{74.25.Nf, 74.70.Ad}

\section{Introduction}
Magnesium diboride ($MgB_2$) is an intermetallic compound whose
superconductivity
with $T_c$ $\approx$ 39 K was recently discovered.\cite {1} The observation
of a significant boron
isotope effect\cite {2} strongly suggests that the system is a phonon-mediated BCS 
superconductor thus making this intermetallic compound very interesting
for its
remarkably high $T_c$ among conventional BCS superconductors.

$MgB_2$ crystallizes in the hexagonal $AlB_2$ type structure, which
consists of alternating
hexagonal layers of $Mg$ atoms and graphite-like honeycomb layers of
$B$ atoms. Powder
samples of isotopically enriched $Mg^{11}B_2$
were prepared with the method described in Ref. [2].
X-ray powder diffraction
measurements confirmed the hexagonal unit cell of $MgB_2$.\cite {1,2}
Magnetization
measurements done at $H$ = 25 Oe  yield a transition temperature
$T_c$ =  39.2 K   with a
shielding volume fraction close to 100\%.\cite {2,3} We have investigated
two different
samples from different batches of polycrystalline enriched  $Mg^{11}B_2$ in
order to check the
reproducibility of the data. The two samples are referred to as
sample 1 and sample 2.
Both samples are in powder form with sample 1 having been ground to a
smaller grain
size. No substantial differences were observed in the NMR
measurements on the two
samples. The onset of superconductivity in sample 2 was determined by
monitoring the
detuning of the NMR circuit occurring at $T_{irr}$ (see Fig. 1). This type
of measurement
corresponds to probing the temperature dependence of the radio frequency
surface
resistance. Thus, as the magnetic field is increased the transition
region broadens due to
the dissipation associated to flux lines motion below $T_c$ and at
7.2 Tesla no detuning can
be observed although the magnetization measurements indicate a
$T_c$ = 23 K at 7 Tesla.
The temperature at which the detuning is first observed is plotted
in the inset of Fig. 1 for
the different magnetic fields and compared to the $T_{irr}$ values
obtained in Ref. [4] from
magnetization measurements. From the good agreement between the two
sets of data one can
deduce that the onset of detuning in the NMR circuit occurs at the
irreversibility line.

\section{NMR Results}
$^{11}B$ NMR and relaxation measurements were performed with
a pulse Fourier transform
spectrometer in an external magnetic field of 7.2 T and 1.4 T. Fig. 2
shows the $^{11}B$
spectrum obtained from the Fourier transform of half of the solid echo
following a $(\pi/2)_0$-$(\pi/2)_{90}$ pulse sequence. In order to cover
the whole spectrum three separate spectra were
recorded at resonance frequencies centered at the three lines and added
together. The
spectrum shows the typical powder pattern for a nuclear spin $I$ = 3/2
in presence of first
order quadrupole effects with an axially symmetric field gradient.\cite {5}
The separation of
the symmetric satellite lines is given by
$\delta \nu = \nu_Q (3\cos^2 \theta - 1)$ whereby for the powder
singularities $\theta$ = 90$^\circ$ thus yielding
$\nu_Q = e^2qQ/2h = 835 \pm 5$ kHz. The axially symmetric
field gradient is consistent with the local hexagonal symmetry at the
boron site. The
quadrupole coupling constant is practically independent of temperature
(see inset in
Fig. 2) ruling out any structural distortion of the lattice in the
temperature range investigated.

The full width at half maximum (FWHM) of the central line of the
$^{11}B$ spectrum is
about 16 kHz in the 7.2 T magnetic field and is temperature independent
in the 4.2 - 300 K
range. The second order quadrupole broadening of the central line at
this high field can
be estimated to be of the order of $\nu_Q^2 / 2\nu_L$ = 3.5 kHz and can
thus be neglected. At 1.4
Tesla the contribution due to quadrupole broadening is 18 kHz. This is
consistent with
the observed $^{11}B$  NMR spectrum which dispalys an asymmetric
central line as expected
for a powder broadening due to second order quadrupole effects.

The $^{11}B$ NMR shift of the central line was measured with respect
to the resonance
frequency in $H_3BO_3$ aqueous solution at 7.2 T.  Since the central line
does not show at
high field  any detectable asymmetry the anisotropic shift must be
negligible and the
measured shift can be entirely ascribed to an isotropic Knight Shift ($K$).
The temperature
dependence of $K$ is shown in Fig. 3. The large error bars are due to
the small size of the shift
with respect to the width of the line.

$^{11}B$ nuclear spin lattice relaxation rates (NSLR) were measured by
saturating the central
line only with a single $\pi/2$  pulse and monitoring the growth of the
solid echo at variable
delays.\cite {6} The NSLR defined as $2W = T_1^{-1}$ was extracted
from the fit of the data to the
recovery law:
$$M(\infty) - M(0)/M(\infty) = 0.1 \exp (-2Wt) + 0.9 \exp (-12Wt)$$
which is obtained from the solution of the master equations in case of
a magnetic
relaxation mechanism.\cite {7} The temperature dependence of the NSLR is
shown in Fig. 4.
As can be seen a linear temperature dependence is observed  in the
normal phase that is
independent of the applied magnetic field. Below  $T_c$  the NSLR is
very different for the
two field values i.e. 7.2 T and 1.4 T.

\section{Discussion and Conclusions}

\subsection{Normal State}
The Knight Shift at the $^{11}B$ site is dominated by the contact
hyperfine interaction due to
$s$-type electrons with negligible contributions from core polarization,
exchange
interactions and orbital terms.\cite {8,9} Thus, the very small value
of $K$ at room temperature
(see Fig. 3) indicates a very small $s$-character of the wave function
of the conduction
electrons at the Fermi surface in $MgB_2$. For comparison, in
$YNi_2B_2C$ intermetallic
compound, where the isotropic Knight Shift is almost one order of
magnitude bigger than
in $MgB_2$,\cite {9} the ratio between the $s$-type density of states
at the Fermi level with respect
to the total density of state was estimated to be
$N_{BS}(E_F)/N(E_F)$ = 0.013. Thus for $MgB_2$
one expects an even smaller ratio. Although the absolute value of such
a small shift
depends critically on the reference compound chosen we can conclude
that the $s$-type
wave function at the Fermi level is negligible in $MgB_2$. This
conclusion is not
inconsistent with the results of band structure calculations.\cite {10}

In the normal state the NSLR follows a linear temperature dependence
down to $T_c$ with
$T_1T$ = 165 $\pm$ 10 sec K  (see Fig. 4). The Korringa
ratio\cite {8} at room temperature is
$K^2T_1T/S \approx$ 0.2, much lower than the ideal value of unity
for $s$-electrons and is  temperature
dependent ($S = (\gamma_e/\gamma_n)^2(h/8\pi^2k_B)$ = 2.57 10$^{-6}$
for $^{11}B$ with $\gamma_e$, $\gamma_n$ the gyromagnetic ratios for
the electron and nucleus respectively).  This indicates that the
NSLR is driven by a
mechanism different from scattering with $s$-type conduction electrons
at the Fermi level.
One possible explanation is that the NSLR is dominated by orbital
contribution due to $p$-electrons. This contribution is expected
to be proportional to the density of p-states at the
Fermi level and to temperature but does not obey the Korringa ratio.\cite {8}

\subsection{Superconducting State}
In conventional BCS $s$-wave superconductors one expects
a coherence peak below $T_c$ 
in the NSLR and an exponential decrease at low $T$ of both the Knight
Shift and $T_1T$ due
to the opening of the superconducting gap.\cite {11} Since the
Knight Shift is small compared
to the line width in a powder sample no quantitative information can be
obtained here
from the decrease of $K$ below $T_c$ (see Fig. 3). Regarding the NSLR
the measurements
in an external magnetic field are difficult to interpret because of
the presence of a
contribution to relaxation due to the flux lines.\cite {12} Also, the size
of the coherence peak
can be drastically reduced by an external magnetic field,
through a pair-breaking mechanism.\cite {13}

At 7.2 Tesla the temperature dependence of $T_1^{-1}$ in Fig. 4 does not
show any coherence
peak nor exponential drop but rather a slight enhancement with respect
to the normal
state which persists down to 4.2 K.  At such a high field  it is likely
that the NSLR below
$T_c$ is largely dominated by the relaxation of nuclei inside
the vortex core. In fact, from
the measurement of the upper critical field $H_{c2}(0)$ = 16.5 T\cite {14}
one can estimate a
vortex size $\xi$ = 60 $\AA$ which can be compared with the intervortex
spacing at 7.2 T for a
triangular vortex lattice i.e. $d$ = 200 $\AA$. In presence of spin
diffusion process (short $T_2$ and
long $T_1$) the measured NSLR is a weighted average of the relaxation
in the vortex core
where the metal is in the normal state and in the intervortex region
whereby the short
relaxation in the vortex core dominates the measured NSLR.\cite {12}
As a matter of fact the
slight increase of the NSLR observed below $T_c$ at 7.2 T is similar to
the effect due to
thermal motion of flux lines previuosly observed in HTSC.\cite {15}

At 1.4 Tesla the ${11}B$ NSLR (see Fig. 4) in the normal state is the
same as at 7.2 T as
expected for a relaxation mechanism due to scattering with the conduction
electrons.
Below the superconducting transition temperature the signal becomes
very weak due to
the small penetration of the radio frequency. The few measurements we
could perform
indicate that there is no measurable coherence peak just below $T_c$ and
that the  NSLR at
low temperatures  is much longer than at 7.2 T confirming that the
dominant contribution
to relaxation in an external magnetic field is associated with the
density of flux lines and
their thermal fluctuations.\cite {12,15}

In conclusion, we compare our results with the NSLR data reported by
Kotegawa et al.\cite {16} Above $T_c$ both sets of data are in good
agreement. Below $T_c$ our data at 7.2 T and
at 1.4T are quite different from Kotegawa's data taken at 4.42 T
confirming the strong
field dependence of $1/T_1$  in the superconducting phase that we
ascribe to the contribution
of fluxons. As a consequence, any conclusion regarding the coherence
peak and the
superconducting gap obtained from data taken in an external magnetic
field appears
dubious. One should either perform the measurements in zero field by a
field cycling
technique\cite {11} or one should do a careful study of the field
dependence in order to
subtract off first the contribution due to flux lines. These studies,
which require a long
time to perform due to the weakness of the $^{11}B$ signal and the
very long NSLR at low
fields, are currently planned. Finally we compare our results with the
results reported by
Gerashenko et al.\cite {17} The conclusions about the $^{11}B$ quadrupole
perturbed NMR
spectrum are in excellent agreement with our data. The differences
in the Knight Shift
values should be ascribed to the use of a difference reference sample.
The data of NSLR
below $T_c$ were taken by Gerashenko et al.  in a magnetic field of
2.113 T and are in
qualitative agreement with our data at 1.4T particularly
regarding the absence of a
coherence peak. 

\section{Acknowledgments}
We would like to thank D. K. Finnemore and V. P. Antropov for useful
discussions and
D. Procissi and R. Vincent for experimental help. Ames Laboratory is
operated for US
Department of Energy by Iowa State University under contract
No. W-7405-Eng-82.
This work at Ames Laboratory was supported by the Director for
Energy Research,
Office of Basic Energy Science.


\begin{figure}
\epsfxsize=0.9\hsize
\vbox{
\centerline{
\epsffile{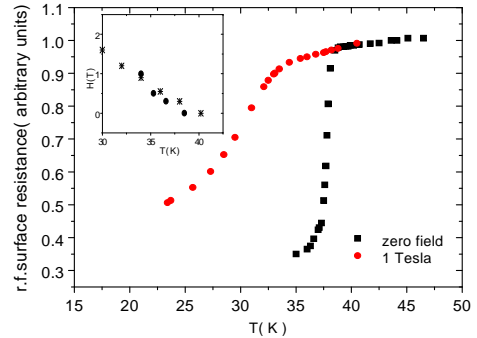}
}}
\caption{Detuning curve of the NMR tank circuit at 100 MHz for $MgB_2$
sample 2 in zero field and at 1 Tesla. The inset shows the temperature
at which detuning is first observed ($\bullet$) and the $H_{irr}$
from Ref. [4] ($\ast$).}
\label{F1}
\end{figure}
\begin{figure}
\epsfxsize=0.9\hsize
\vbox{
\centerline{
\epsffile{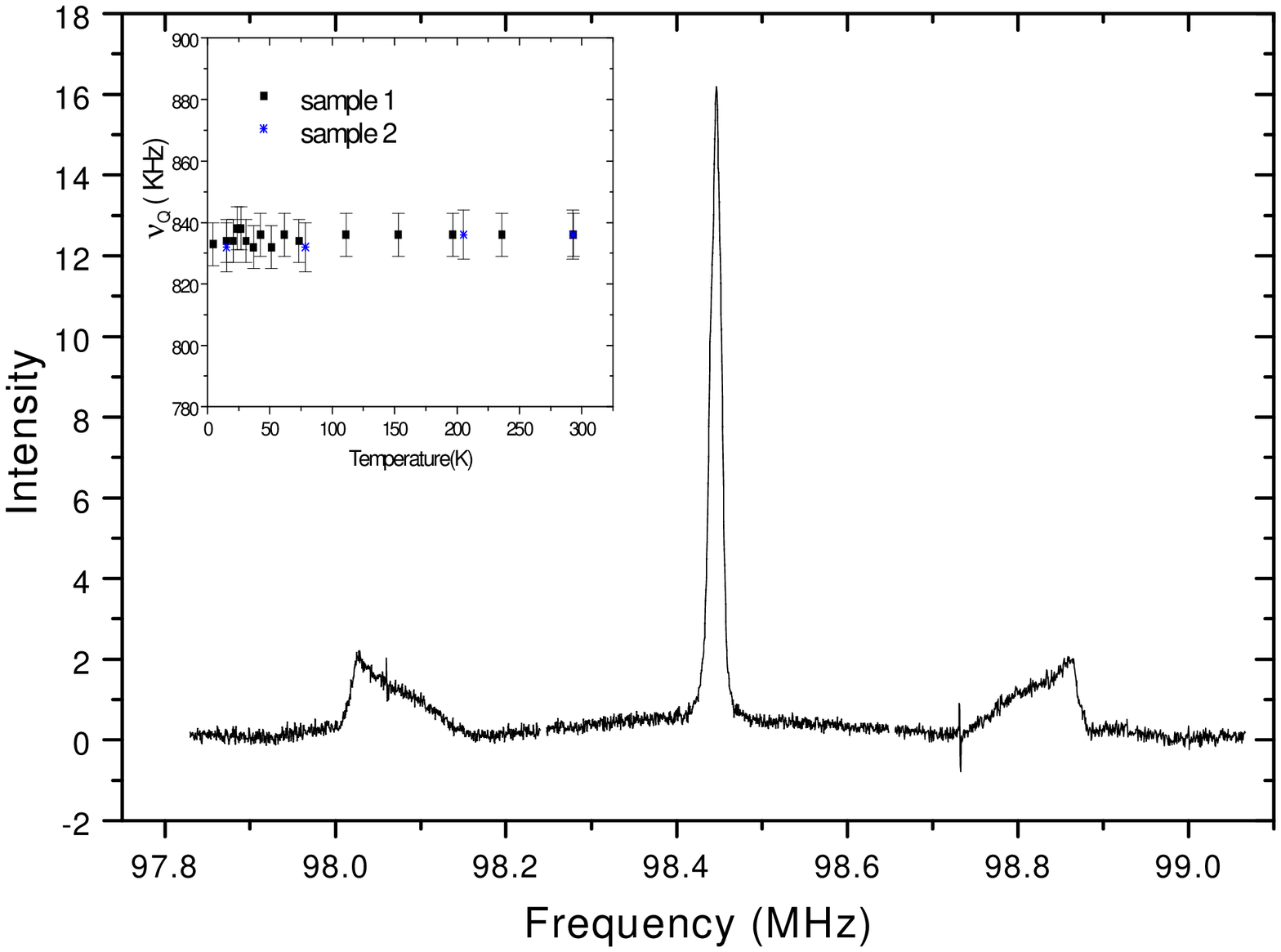}
}}
\caption{Room temperature $^{11}B$ NMR spectrum for $MgB_2$ powder sample 1
showing both the central line transition and the two singularities of
the distribution of satellite transitions. The inset shows the temperature
dependence of the quadrupole coupling constant derived from the spectrum.}
\label{F2}
\end{figure}
\begin{figure}
\epsfxsize=0.9\hsize
\vbox{
\centerline{
\epsffile{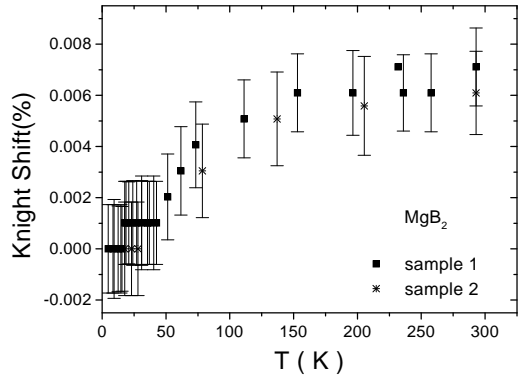}
}}
\caption{Temperature dependence of the Knight shift in $MgB_2$, sample 1
and sample 2. The shift is with respect to the $^{11}B$ resonance
in a water solution of boric acid at room temperature.}
\label{F3}
\end{figure}
\begin{figure}
\epsfxsize=0.9\hsize
\vbox{
\centerline{
\epsffile{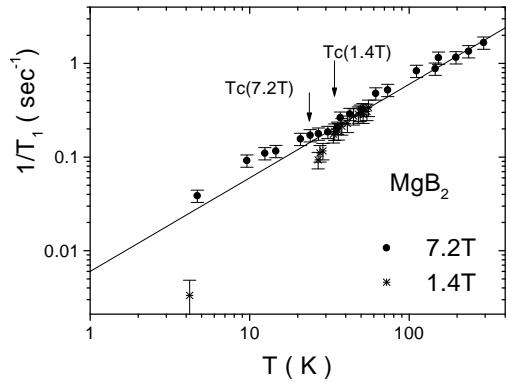}
}}
\caption{Temperature dependence of $^{11}B$ NSLR in $MgB_2$  at 7.2 Tesla
and at 1.4 T.  The line is the Korringa law  with $T_1T$ = 165 sec K.
The arrows indicate the superconducting transition temperature at
the two fields.}
\label{F4}
\end{figure}

\vfil\eject

\end{document}